\documentclass{elsarticle}
\usepackage{graphicx}
\usepackage[usenames]{color}
\begin{document}
\begin{frontmatter}
\title{Ironsilicide formation by high temperature codeposition of Fe-2Si with different thicknesses on Si (111)}
\author[Budapest]{I. D\'{e}zsi\corref{cor1}\fnref{*}}
\ead{dezsi@kfki.rmki.hu}
\author[Budapest]{Cs. Fetzer}
\author [Budapest]{F. Tanczik\'{o}}
\address[Budapest]{KFKI Research Institute for Particle and Nuclear
Physics H-1525 Budapest 114 P.O.Box 49, Hungary}
\author[Budapest2]{P. B. Barna}
\author[Budapest2]{O. Geszti}
\author[Budapest2]{G. S\'{a}fr\'{a}n}
\author[Budapest2]{L. Sz\'{e}kely}
\address[Budapest2]{Research Institute for Technical Physics and Materials Science
Budapest 114 P.O. Box 49, Hungary}
\author[Leuven]{H. Bender}
\address[Leuven]{IMEC Kapeldref 75, B-3001, Leuven, Belgium}
\cortext[cor1]{ Corresponding author}
\begin{abstract}
Fe and 2Si were co-deposited on Si (111) surface at 853 K. The formation of
silicides was investigated by M�ossbauer spectroscopy and electron microscopy.
Depending on the thickness of the deposited films different phases were formed.
At low thickness, stable $\varepsilon$-FeSi (B20) and metastable [CsCl]-Fe$_{1-x}$Si (B2) phases were observed. In the latter case,
because of the presence of Fe vacancies in the lattice the local symmetry around the iron components was lower than cubic.
At larger (12 nm) deposited thickness, stable $\beta$-FeSi$_2$ has been formed.
\end{abstract}
\end{frontmatter}
\section{Introduction}
The formation of iron silicide phases on silicon substrate attracted
considerable attention in the last decades
\cite{Calandra,Murarka,Bost,Desimoni,Borisenko,Comrie}. The
particular aim was to determine the Fe-silicide structures and
possibly to grow epitaxial semiconducting $\beta$-FeSi$_2$ layers
on/in Si, since $\beta$-FeSi$_2$ has a direct gap of 0.87 eV, and
could be used in  optoelectronic device fabrication.  However, the
formation of iron-silicides proved complex processes, especially, depending on
the preparation methods. The interface formation involves  reactive
intermixing between iron and silicon at the deposition process and
the resulting Fe-silicide forms on the surface of the Si substrate.
First,  Fe depositions were made at room temperature,  later on
heated Si substrate \cite{Derien}. In order to decrease possibly the
out diffusion of the substrate Si atoms  during the formation of the
Fe-silicide phases on the surface, the films were grown by co-depositing Fe and Si in different \cite{Kane,Wang} concentration ratio (1:1, 1:2). The
experiments resulted in various iron silicide phases. The phases crystallized in cubic structure \cite{Kane,Onda} and in non-stoichiometric composition. The structure and morphology aspect of the Fe-silicide  was investigated \cite{Ji} by X-ray diffraction (XRD) and reflective high energy electron diffraction (RHEED) methods during the  co-deposition of Fe and Si in 1:2 ratio on Si(111) during
different  deposition times at 580$^\circ$C.  It was observed that during the growth process, first separate then intergraded islands (in layer form)  were formed. The RHEED images were observed during the depositions at low  thickness and  the images were interpreted by the formation of $\gamma$-FeSi$_2$. At thicker depositions, the $\beta$-FeSi$_2$ phase was observed.
\newline In order to gain more information on the Fe-silicide phases forming at different thicknesses  we performed M\"{o}ssbauer and electron spectroscopy
studies on the co-deposited Fe-2Si samples in order to get  correct information
on the  different Fe-silicide phases, especially, on the formation of the $\gamma$-FeSi$_2$   phase. The M\"{o}ssbauer effect
using conversion electron spectroscopy (CEMS) is a very sensitive
method to observe  correct phase formation on  surfaces between iron and silicon.
The electron microscopy, especially, in transition electron microscopy (TEM) mode is very effective for structural determination of the phases. The parallel application of the two methods may provide reliable results on the formation of different phases depending on the different thicknesses.

\section{Experimental Details}

High purity Si (111) substrates were chemically etched in an aqueous
solution containing 10 mol. percent HF and 40 mol. percent NH$_4$F
to get rid of the oxide on the surface. In a pre-annealing chamber
($\leq$ 1 x 10$^{-10}$ mbar), the samples were introduced and heated
to 875 K for 30 min. After this treatment, the Si substrates were
transported into the chamber of an MBE (MECA) system. Fe
and Si atoms were co-deposited with 0.052 : 0.177 volume ratio at 853
K, in 1Fe : 2Si atomic ratio.  The base pressure in the chambers was
1x10$^{-10}$ mbar and increased to 2x10$^{-9}$ mbar during the
depositions. $^{57}$Fe was evaporated by using Knudsen-cell with BeO
crucible. Si was evaporated by electron gun. The deposition rate was
0.117 {\AA}/sec. Four layers of FeSi$_2$ with 12, 5, 2.5 and 1,3 nm
thicknesses were prepared.
\newline
M\"{o}ssbauer measurements were carried out by using a conventional
constant acceleration-type spectrometer. For the detection of the
conversion electrons a low-background proportional counter filled
with He and 4 vol. percent of CH$_4$ was used at room temperature.
For the analysis of the spectra, a least-squares fitting program was
used. Also, using this program, spectra with histogram distributions
of the parameter values could be fitted. The isomer shift values are
given relative to that of $\alpha$-Fe at room temperature.\newline
Specimens for in-plane electro-microscopic measurements were thinned
by ion milling of the back side. The incidence angle of the iron
beam was $\sim$ 3-5$^\circ$ as measured from the surface plan of the
specimen.   The conventional TEM investigation of the specimens was
carried out in a PHILIPS CM20 transmission electron microscope
equipped with an EDX analyzer. The high resolution TEM investigation
was carried out in a 300 kV JEOL 3010 TEM and an attached GATAN
TRIDIEM Image FILTER was used for elemental mapping.
\section{Results}
The  M\"{o}ssbauer spectra of the  samples  measured at room
temperature are shown in Fig. 1.

\begin{figure}
\begin{center}
\includegraphics[width=8cm]{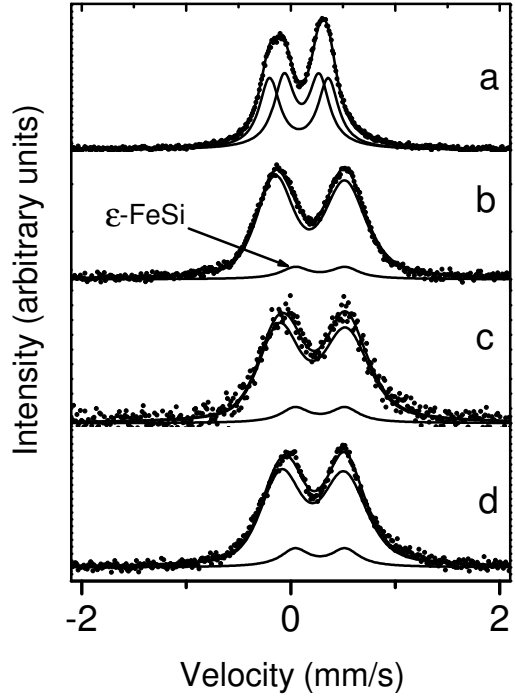}
\caption{M\"{o}ssbauer spectra of deposited FeSi$_2$ layers. Thicknesses of the layers given in nm a: $\beta$-FeSi$_2$, b: 5, c: 2.5, d: 1.3.}
\label{Fig1}
\end{center}
\end{figure}

Quadrupole split  spectra were
observed for all samples with 12, 5, 2.5 and 1.3 nm  thickness values.
In the spectrum of 12 nm thick  sample (Fig. 1a), two
quadrupole split doublet spectra of Lorentian shape could be fitted
with the hyperfine parameter values characteristic to the
$\beta$-FeSi$_2$ phase \cite{Fanciulli}. Spectra of 5, 2.5 and 1.3 nm
thick samples showing broad quadrupole split spectra were fitted by parameter distributions  because their
shapes were not Lorentzian and  considerably broadened.  The fixed parameters of $\varepsilon$-FeSi
phase were also included (see later). The average
isomer shift ($\delta$),  average quadrupole spliting (QS)  values are compiled in Table I.
\begin{table}[tpb]
\caption{M\"{o}ssbauer parameters of codeposited Fe-2Si samples determined at room temperature.  Sample thickness in nm . Isomer shift ($\delta $), quadrupole
splitting (QS) and width (W) values are
given in mm/s. Relative intensities  (R. int) are given in percent. Average values are denoted by ${< >}$.}\centering
\par
\begin{tabular}{cccccc}
Thickness nm&Component& $\delta$ & QS&  W&R. int\\
\hline
12  &1& 0.08(2) & 0.56(2)&  0.24(1)&50(2)\\
           &2& 0.10(2) & -0.34(1)& 0.24(1)&51(2)\\
5  &1& 0.18 & 0.66 & 0.26& 9.2(2)\\
           &2&$ <$0.28$>$&$<$0.67$>$&0.26 & 90.8(2)\\

2.5&1& 0.18 & 0.66 & 0.26& 11.2(2)\\
&2&$<$0.21$>$&$<$0.64$>$  &0.26 & 11.2 (2)\\
1.3
&1&0.18&0.66 &0.26 & 13.5(3)\\
&2&$<$0.21$>$&$<$0.61$>$ &0.26 & 86.5(1)\\
\end{tabular}
\end{table}

The structure of the 12 and 5 nm  thick samples  has been investigated by  transmission  electron microscopy and selected area  electron diffraction (SAED) in details both on plane view  and cross sectional specimens.
\begin{figure}
\begin{center}
\includegraphics[width=8cm]{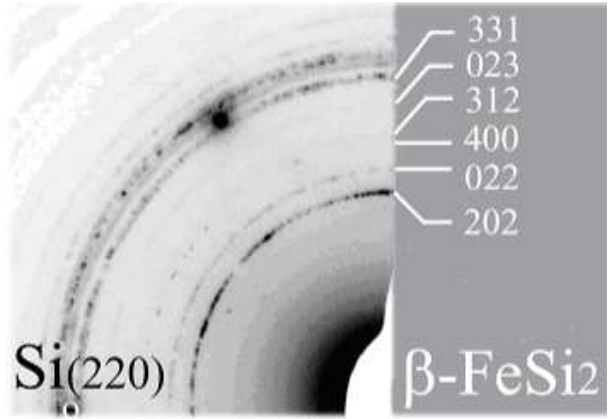}
\caption{Selected area electron diffraction pattern of the plane view specimen of the 12 nm thick sample.}
\label{Fig2}
\end{center}
\end{figure}

 The SAED pattern  clearly show that the $\beta$-FeSi2 crystals  are randomly oriented  and no epitaxial  growth existed. Cross sectional TEM images of the 12 nm sample (Fig. 3) clearly show that the Fe-Si layer is continuous and polycrystalline. In Fig. 3a, b and c $\beta$-FeSi$_2$ (311) (0.28 nm) lattice fringes are shown. Grain boundaries are shown by arrows in Fig. 3a.
\begin{figure}
\begin{center}
\includegraphics[width=8cm]{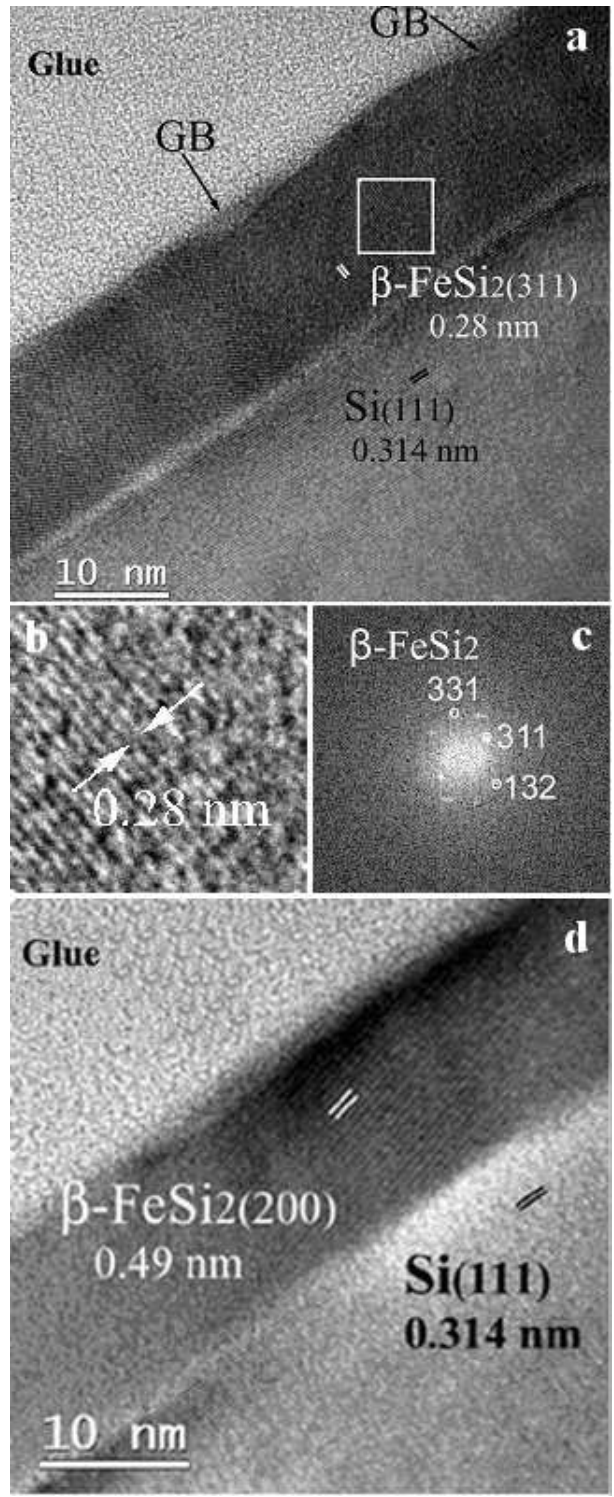}
\caption{XTEM images of 12 nm thick sample. (a) overview, (b)  the 311 lattice fringes (0.28 nm of $\beta$-FeSi$_2$ phase (c) corresponding FFT patterns, (d) 200 lattice fringes (0.49 nm) of $\beta$-FeSi$_2$.}
\label{Fig3}
\end{center}
\end{figure}

The plain-view bright field (BF) TEM image and the corresponding  elemental Fe EELS map of the 5 nm  thick sample are shown in Fig. 4a and b.
\begin{figure}
\begin{center}
\includegraphics[width=8cm]{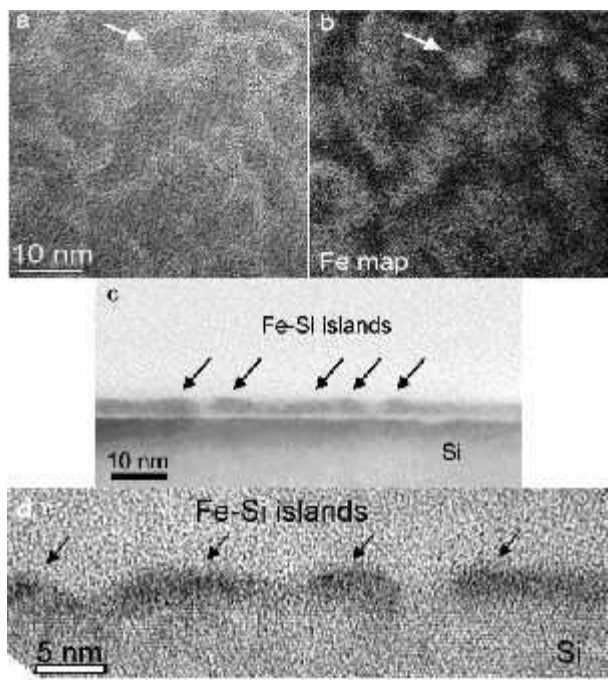}
\caption{TEM images of the 5 nm thick sample. (a) in plane bright field image, (b) corresponding Fe elemental map showing the Fe-Si islands on the Si substrate. The arrows indicate one of the islands. (c) and (d) cross section of the sample.}
\label{Fig4}
\end{center}
\end{figure}

These images clearly show the island structure of the film  on the Si (111) substrate.  The corresponding TEM  BF image  and EELS elemental map represent the same area  (one of islands is marked  by arrows on both images). In the bright field image, the Fe/Si islands show dark contrast due to the higher atomic number of Fe, while in the Fe elemental EELS map the Fe/Si islands are shown as  bright contrast. Fig.4c and d demonstrate also  the domain structure  of the Fe la-Si  layer in cross section. The SAED analysis of the crystal structure is shown in Fig.5.
\begin{figure}
\begin{center}
\includegraphics[width=8cm]{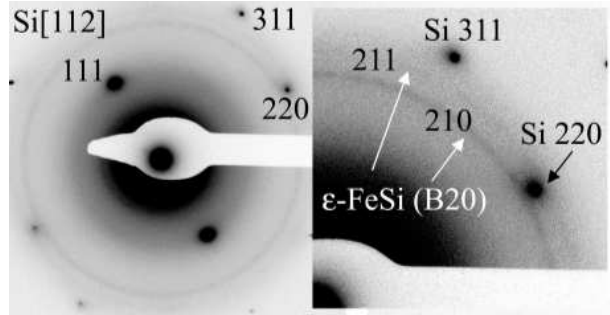}
\caption{SAED pattern of the plane view specimen of the 5 nm sample.
 }
\label{Fig5}
\end{center}
\end{figure}

In this diffraction pattern  only the $\varepsilon$-FeSi (B2) [ JCPDS 76-1748 ] phase can be identified. Cross sectional TEM images of the 5 nm thick sample is shown in Fig. 6.
\begin{figure}
\begin{center}
\includegraphics[width=8cm]{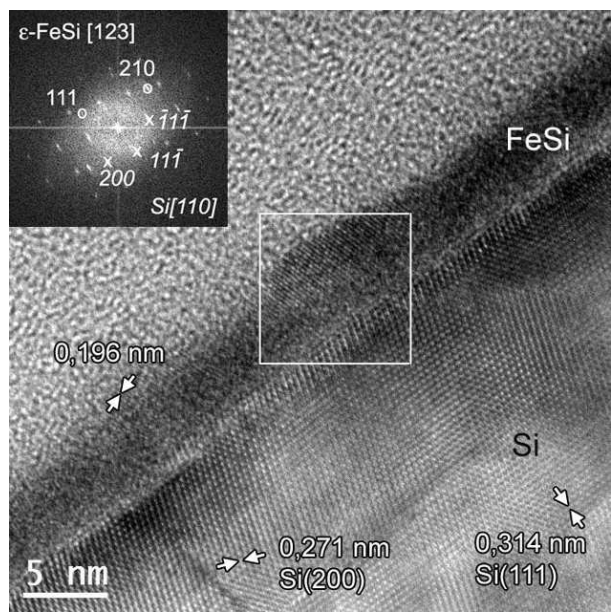}
\caption{Cross section HRTEM image of the 5 nm thick sample. }
\label{Fig6}
\end{center}
\end{figure}

Lattice spacings of  0.190-0.199 nm  as single set of lattices have been found in several islands. These could belong to the (210) lattice planes  (0. 199 nm) of the $\varepsilon$-FeSi identified  by the SAED analysis of the plane view specimen. In Fig. 7, the high
resolution TEM image is shown in the Si [112] direction.

\begin{figure}
\begin{center}
\includegraphics[width=8cm]{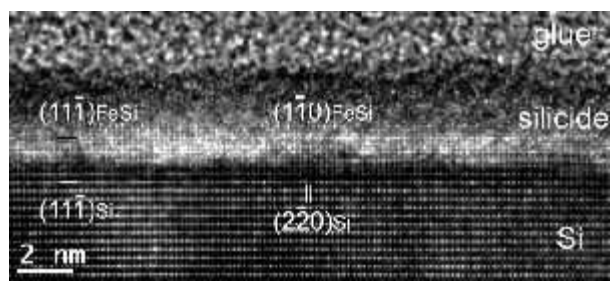}
\caption{HRTEM image of the Si/Fe-Si interface taken along the Si [112] zone. The phase in contact with the single crystal Si could be identified as [CsCl]-FeSi (B2).}
\label{Fig7}
\end{center}
\end{figure}

The layer
is grown epitaxially on the Si substrate surface and  it could be
identified as [CsCl]-Fe$_{1-x}$Si (B2) [JCPDS26-1141].
The spectra of the thinner deposits showed thinner and smaller islands
but the same images as the 5 nm thick sample.
\section{Discussion}
The HRTEM and SAED as well as FFT and M\"{o}ssbauer analysis of the
12 nm thick Fe-Si sample proved the formation of  $\beta$-FeSi$_2$  in
the co-deposited layer. The film  was continuous with randomly
oriented grains corresponding to the structure evolution of the
poly-crystalline film. The  $\beta$-FeSi$_2$ phase formed because of
the relatively large formation heat  and the lower effect of the Si substrate surface potential for the thicker layer. No other phases could be detected.\newline
The structural results of the thinner layers appeared more complex.
The morphological evolution during the thin film growth is
a complex process  resulting  from the inhibition on the time-scale
of deposition and to reach an equilibrium state. The depletion zones
can have different dimensions depending on the substrate structure
and the surface energy values. The high temperature mainly results
in local rearrangement of the atoms, not in long range diffusion or
inducing chemical reaction with surface atoms
as it is in the case of reactive deposition to the surface of a substrate.
The formation of phases depend also on the
heat of formation of the different phases \cite{Pre,Pre2}.
Concerning the Si - Fe phases, the $\varepsilon$-FeSi has the maximum
heat formation value calculated in \cite{Mie}.
$\varepsilon$-FeSi has cubic structure but the M\"{o}ssbauer
spectrum of $\varepsilon$-FeSi shows quadrupole split spectrum
because of the lower than cubic symmetry in the nearest neighborhood
of the Fe in the lattice. This doublet may overlap with the other doublet
component(s) in the broadened  spectrum of 5 nm  and in the thinner samples.  The broad spectrum could be fitted by fixing the known hyperfine parameter
values \cite{Fanciulli2}. For $\varepsilon$-FeSi,  9(2) percent relative
intensity value was obtained  for the 5 nm thick sample.   The relative
intensities for this phase are somewhat larger in thinner samples.  Since the spectra are  overlapped with the broad spectrum the exact real line width of the $\varepsilon$-FeSi component can not  be exactly known and  this may effect the relative intensity of this component. The broad quadrupole doublet can be interpreted by the formation of disordered
[CsCl]-Fe$_{1-x}$Si as it is mentioned in Results caption. The formation of [CsCl]-Fe$_{1-x}$Si  can be the result of the stabilization of the Fe-Si positions by forming epitaxial
structure on the Si surface and  in the positions where  small
Si crystal is present in the deposited layer. The high resolution TEM image shows
that the [CsCl]-Fe$_{1-x}$Si (B2) phase is present
in the domain islands on the Si surface. This phase was also observed
after laser treatment of $\varepsilon$-FeSi layer on  Si substrate
\cite{Falepin}.  The hyperfine interaction values of the thin layers having parameter distributions (Table I.) and  are in the range of the values published by the referred authors.
The illustration of  the Fe$_{1-x}$Si lattice with Fe vacancies is shown in Fig. 8.   \begin{figure}
\begin{center}
\includegraphics[width=9cm]{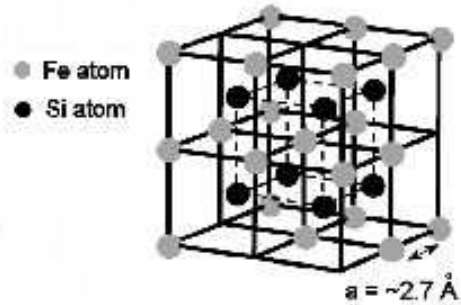}
\caption{The lattice structure of non-stoichiometric [CsCl]-Fe$_{1-x}$Si.}
\label{Fig8}
\end{center}
\end{figure}

The structure of the $\gamma$-FeSi$_2$ phase has  cubic fluorite-type structure \cite{Derien2}. This case, the M\"{o}ssbauer spectrum should  appear as a  single line. But such component did not appear in our spectra in agrement  with  the measured TEM image. The reason is that the structure contains Fe vacancies positioned not regularly resulting in electric field gradient at the Fe sites, consequently, quadrupole split M�ssbauer spectra with the distribution of the parameter values. At higher deposited thickness of Fe-2Si,  $\varepsilon$-FeSi and [CsCl]-Fe$_{1-x}$Si phases do not form.

\section{Conclusions}
The formation of different  phases are determined  after co-deposited  Fe-2Si
on Si (111). It is clearly shown that  the formation of phases  depends on the
thickness of the co-deposited layer on the surface at 853
K. At relatively low deposited thickness  two phases form.  One is the $\varepsilon$-FeSi stable
phase. The other  is [CsCl]-Fe$_{1-x}$Si  with the hyperfine parameter values
characteristic to a disordered arrangement around the lattice sites of Fe. The metastable $\gamma$-FeSi$_2$ phase did not form in relatively thin  samples. In the
12 nm thick sample, pure stable poly-crystalline $\beta$-FeSi$_2$  has been formed.
\section{Acknowledgement}
This work was supported  by the Hungarian National Research Fund (OTKA) project No. K62272 .

\end{document}